\documentclass[preprint,aps]{revtex4}
\usepackage{graphicx}
\usepackage{booktabs}
\usepackage[usenames, dvipsnames]{color}
\usepackage{multirow}
\usepackage{tabularx}
\usepackage{setspace}
\usepackage{url}
\usepackage{array,booktabs}
\usepackage{rotating}
\usepackage[section]{placeins}
\usepackage[ pdftex, plainpages = false, pdfpagelabels, 
                 pdfpagelayout = useoutlines,
                 bookmarks,
                 bookmarksopen = true,
                 bookmarksnumbered = true,
                 breaklinks = true,
                 linktocpage=all,
                 pagebackref=false,
                 colorlinks = true,
                 linkcolor = BrickRed,
                 urlcolor  = blue,
                 citecolor = BrickRed,
                 anchorcolor = green,
                 hyperindex = true,
                 hyperfigures
                 ]{hyperref} 
                 
\newcommand{\comment}[1]{}

\newcolumntype{L}[1]{>{\raggedright\arraybackslash}p{#1}}
\newcolumntype{C}[1]{>{\centering\arraybackslash}p{#1}}
\newcolumntype{R}[1]{>{\raggedleft\arraybackslash}p{#1}}

\makeatletter
\def\url@leostyle{%
  \@ifundefined{selectfont}{\def\UrlFont{\sf}}{\def\UrlFont{\small\ttfamily}}}
\makeatother
\begin{document}

\title{\href{http://www.necsi.edu/research/social/}{Risking It All: \\
Why are public health authorities not concerned about \\
Ebola in the US? \\
Part I. Fat tailed distributions}} 
\author{\href{http://necsi.edu/faculty/bar-yam.html}{Yaneer Bar-Yam}}
\affiliation{\href{http://www.necsi.edu}{New England Complex Systems Institute}\\  
210 Broadway, Suite 101, Cambridge MA 02139, USA}
\date{Nov. 5, 2014}

\begin{abstract}
US public health authorities claim imposing quarantines on healthcare workers returning from West Africa is incorrect according to science. Their positions rely upon a set of studies and experience about outbreaks and transmission mechanisms in Africa as well as assumptions about what those studies imply about outbreaks in the US. According to this view the probability of a single infection is  low and that of a major outbreak is non-existent. In a series of brief reports we will provide insight into why properties of networks of contagion that are not considered in traditional statistics suggest that risks are higher than those assumptions suggest. We begin with the difference between thin and fat tailed distributions applied to the number of infected individuals that can arise from a single one. Traditional epidemiological models consider the contagion process as described by $R_0$, the average number of new infected individuals arising from a single case. However, in a complex interdependent society it is possible for the actual number due to a single individual to dramatically differ from the average number, with severe consequences for the ability to contain an outbreak when it is just beginning. Our analysis raises doubts about the scientific validity of policy recommendations of public health authorities. We also point out that existing CDC public health policies and actions are inconsistent with their claims.
\end{abstract}

\maketitle

The Ebola epidemic became a global public health crisis due to its persistent exponential growth over half a year despite efforts to contain it in West Africa.
The arrival of the first Ebola case by commercial flight to the US, Thomas Duncan, and subsequent infections in Dallas of two nurses treating him, led to multiple stages of reevaluation of public health policies \cite{weiss,CDC1,CDC2,CDC3,CDC4,DHS}. These changes in policy resulted from unanticipated events and constituted additional precautions in hospital settings and in monitoring of travelers from West Africa to protect the public. A second case, an infected physician returning to New York from providing care in Guinea, triggered a set of new policies. Some of these are not by public health authorities but by elected officials in New York, New Jersey, Illinois and elsewhere \cite{States1,States2,States3}, and include 21 day quarantines or home observation of returning health care workers. The response of public health authorities has been to affirm \cite{Fauci,NEJM} that these policy choices are counter to science and experience from Ebola outbreak response in Africa \cite{Dowell,Roels,Kersteins,Baron,Muyembe,Francesconi,Wamala}. 

A central aspect of the understanding of the risk of outbreaks for Ebola is the mechanism of transmission. Generally, it is understood that those who have been in direct physical contact with individuals with Ebola resulting in exposure to their body fluids, or indirectly with their body fluids on surfaces, have a probability, not a certainty, of being infected. Healthcare workers adopt precautions against infection including personal protective equipment (PPEs) that are essentially hazardous materials suits. Nevertheless, a large number of infections and deaths of healthcare workers have occurred \cite{whoreport}. Different kinds of direct and indirect contact have different levels of risk. Evidence suggests that there is a coincidence between a transition from a non-symptomatic, non-transmissible (latent) period and a period in which observable symptoms start, transmission is possible, and blood tests are positive. Policies are designed, therefore, to assure that individuals are not in contact with others, and PPEs are being used, after symptoms start. Hence the importance of isolation once symptoms start, and monitoring prior to symptoms to ensure that a transition to the symptomatic state does not result in cases of infection.   

Public health authority positions, including a New England Journal of Medicine (NEJM) editorial \cite{NEJM}, 
state there is no need for quarantine because isolation is only needed when symptoms are present. According to this view it is sufficient to monitor and isolate when symptoms arise. The NEJM editorial specifically frames its statement as relevant to healthcare workers and does not explain why they did not adopt this stance for others who have been exposed.  Quarantines were imposed on four members of the family who were being visited by Thomas Duncan, the first US Ebola case in Dallas, when he developed symptoms. None of them had symptoms at the time of the quarantine or subsequently developed them \cite{Duncan1,Duncan2,Duncan3,Duncan4,Duncan5}. Since the properties of transmission apply equally to Duncan's personal contacts as to healthcare workers, is it unclear, why NEJM did not write an editorial about the imposition of quarantines on them nor refer to it in the editorial that was written. An explanation of the discrepancy is of importance as it leaves the public and officials uncertain about how to interpret the inconsistency, and specifically the claim that quarantine does not reflect a valid policy option. Perhaps just as significant, public discussions \cite{Duncan4} imply that the CDC has been advocating voluntary acceptance of home isolation that may not differ from enforced home quarantine, reflecting a legal rather than medical or scientific distinction. These issues suggest that the position that is being taken by public health authorities is not so well defined as a scientific one as they indicate. 

CDC policy for monitoring and restrictions on individuals \cite{CDC4} provides for a number of levels of risk of an individual being infected with Ebola based on exposure and suggests different levels of self-monitoring or health official monitoring as well as restrictions on travel and contact between individuals for each level. For example, public places and distances of less than three feet from others are to be avoided for individuals at high risk \emph{even without symptoms}. The specification indicates enforced restrictions on travel, i.e. ``These individuals are subject to controlled movement which will be enforced by federal public health travel restrictions; travel, if allowed, should occur only by noncommercial conveyances...'' Authorities are given wide latitude to enforce higher levels of restriction based upon their judgements about exposure and likelihood of compliance with voluntary restrictions. Whether or not different levels of self or publicly imposed isolation and monitoring are considered quarantine is a matter of recently adopted terminology. The existence of various levels of risk and associated protocols manifests tradeoffs of risk versus restrictions, and levels of trust in the individuals to self monitor and report their conditions as well as self-restrict activities. The CDC policy identifies returning healthcare workers as being at a medium level of risk (termed ``some risk''). It is important to realize that different categories of risk should have different isolation policies only if there is a societal tolerance for the associated risks, i.e. this is not a question of the existence or absence of transmission but of the associated risk levels. Mathematically, for each risk level there is a probability $I_r$ of an individual being infected in the latent period before it is known. Then there is a probability $T_p$ given a particular imposed protocol that an infected individual will transmit the disease, causing another infection. The product of these two is the probability of an individual in that risk class and following that protocol causing another infection $P_{r,p} = I_r T_p$. If a certain level of risk is acceptable, so that we want the probability of a new case to be lower than a value $P_0$, i.e. $P_{r,p} < P_0$, and if the risk of the individual being infected is lower in one class than another, the protocols can be relaxed allowing $T_p$ to be higher. To prevent an outbreak from growing the number of newly infected individuals only has to be less than one. Acknowledging that we want a significantly lower probability, we can take it to be much less than one, say one in 10, or even one in 100, out of ``an abundance of caution,'' and this would be enough to control an outbreak. We see from this discussion that the essential question is what are the risks that are being taken.

The question of risks is not directly addressed by the public health authorities \cite{NEJM}, except to dismiss ``cynics'' who unreasonably demand $100\%$ certainty. They do not explain what are the risks they consider reasonable and what they do not. Still, one can infer from their statements that the possibility of individual cases is acceptable and that they are concerned only for a large outbreak, whose risk they consider to be non-existent. Under these conditions, the taking of risks might be justified, after all a public fear of Ebola is only justified if it can be the cause of widespread outbreak and not the deaths of a few individuals, of which there are many causes in society. This view has been repeated multiple times, i.e. that the public at large will not be at risk (even if there are a few cases) in the US \cite{presidentatCDC,FriedenSept30}. Indeed, the existence of risk is clearly acknowledged in the risk levels associated with the CDC policies. Low enough levels of risk of a person being a carrier of Ebola can be addressed by lower levels of observation and isolation. If a case arises, then the consequences must not be very severe, otherwise such a risk stratification does not make sense. 

The assessment of risks in CDC policies are composed of a combination of medical knowledge, behavioral assumptions about individuals and risk tolerance. The medical knowledge requires an understanding, among other things, of the coincidence of contagion, symptoms and blood tests; the behavioral assumptions have to do with the levels of trust in such matters as whether instructions are sufficiently clear and people will actually follow them; and the risk tolerance has to do with both the certainty of our knowledge and the levels of risk society is willing to take. An important aspect of any evaluation of these issues is the sensitivity of the policy conclusions to the level of uncertainty in our knowledge, i.e. a sensitivity analysis. The statements of public health authorities are equivalent to the statement that the policies are robust to the uncertainties in our knowledge and statistical variation in transmission and contagion events. 

We will show that this statement is correct only if traditional statistical assumptions are valid. That traditional assumptions are not likely to be applicable to the current conditions leads to a different understanding that is better described by complex systems science \cite{DCS}. Thus, public health authorities are advocating policies that put many at risk. Moreover, their claims that science is behind their positions may discredit science at a time when it is important to demonstrate that science provides clear and effective policy recommendations. Given a higher level of risk, the claims by public health authorities that they do not want to discourage or inappropriately inconvenience healthcare workers who are devoting their service to containing the outbreak is invalid. Healthcare workers devoted to service of the public in containing and preventing the Ebola outbreak should see a conservatively designed isolation protocol after returning from treatment of patients in West Africa as part of that service. This is true whether or not the protocol is called a quarantine. In order to explain these issues, we will provide a set of discussions of particular complex systems concepts and their mathematical basis. In this, the first, we will look at the possibility that a single individual might cause a much larger number of infected individuals than is expected. Even one such case may cause a breakdown in our ability to contain the outbreak.  

As a first step in this discussion we consider the role of assumptions about the distribution of transmission probabilities relevant to standard models of contagion. The ability to monitor sufficiently to prevent large numbers of transmission is at issue. 

Traditional statistical studies are based upon the assumption that probability distributions are ``thin tailed'' (often Gaussian distributions) so that the range of actual values is narrow, confined to a few standard deviations from the mean. In cases where the mean is proportional to the number of elements, any particular instance is not very far away from the average value. Traditional models of contagion assume that every infected individual gives rise to a number $R_0$ of subsequent infected individuals on average. But is this the value we can reliably expect to happen in a particular instance? The answer to this question is of critical importance, especially in the first few stages of an outbreak. 

Complex systems science suggests that distributions are not always thin tailed, they can be fat tailed (also called heavy or long tailed). Among fat tailed distributions are power law (scale free) distributions that are often observed. This means that there are situations in which a single sick individual can give rise to 10s, 100s or 1000s of cases in a single generation. The likelihood of this happening may not be as high as having only a few, but it cannot be dismissed as it can in a thin tailed distribution. 

The difference between thin and fat tailed distributions has been well explained in discussions of risk by Nassim Taleb \cite{BlackSwan}. As an illustration he considers the difference between weight and wealth. Randomly taking $1,000$ people, their total weight is about $1,000$ times the weight of an average individual, but their wealth may be dominated by the wealth of a single one. It is impossible to find a person who weighs 10 times the average weight of an individual (thin tailed distribution). On the other hand a single person can have as much wealth as the two billion poorest people (fat tailed distribution). 

In the Ebola outbreak in West Africa, the value of $R_0$ is about 2 with various studies providing specific estimates \cite{Meltzer,R02,R03}. This means that every infected individual leads to only about 2 other infected individuals. At this value it takes 10 sequential infections (generations) to have 1,000 sick individuals. Since each generation takes a couple of weeks, this is about 5 months which is approximately the time from March, when the outbreak started in earnest, till the end of the summer. As many keep telling us \cite{dismiss1,dismiss2,dismiss3} this still does not make Ebola a major cause of death compared to others like Malaria. If it keeps going, however, the result is disastrous with 1 million sick after 20 generations and 1 billion after 30, in two and a half years. Still, this is different from thousands in a single generation, in which case it would only take 3 generations to get to a billion. A disease that is airborne, e.g. measles, might have an $R_0$ of 10 or more, in which case 9 generations or just over 4 months would be enough to get to a billion sick people.

The probability distribution that naturally describes the number of contacts that are infected by a single individual in a standard contagion model is a Poisson distribution. This distribution describes the variation in number of infection events that happen randomly in a given period of time, i.e. the infectious period. Similar to the Gaussian distribution, the Poisson distribution is a thin tailed distribution. In this distribution the standard deviation is just the square root of the average, i.e. $\sigma=\sqrt{R_0}$, so that the chance of having more than two or three times the typical number is vanishingly small for an outbreak in which $R_0$ is greater than one. This is not the case if there is a long tail distribution.

Figure 1 illustrates the difference between thin and fat tailed distributions, showing tails of the number of possible instances that occur with more than a certain number (cumulative Poisson and power law distributions). So if we think of this as new Ebola cases infected by a single individual, we see that the chance of larger numbers falls much faster for the thin Poisson distribution than for a fat power law distribution. In the power law case we have to identify a level of risk we care about. We see that 3 or more cases will happen 1 in 10 times, 20 or more cases 1 in 100 times, 30 or more cases 1 in 200 times. In the real world the numbers might be higher or lower, but we do not know them from the information available from previous outbreaks. In the Poisson distribution, we do not have to worry about questions like this. Having more than a few cases just would not happen, as the public health authorities have claimed.

\begin{figure}
\centering
\includegraphics[width=0.49\textwidth]{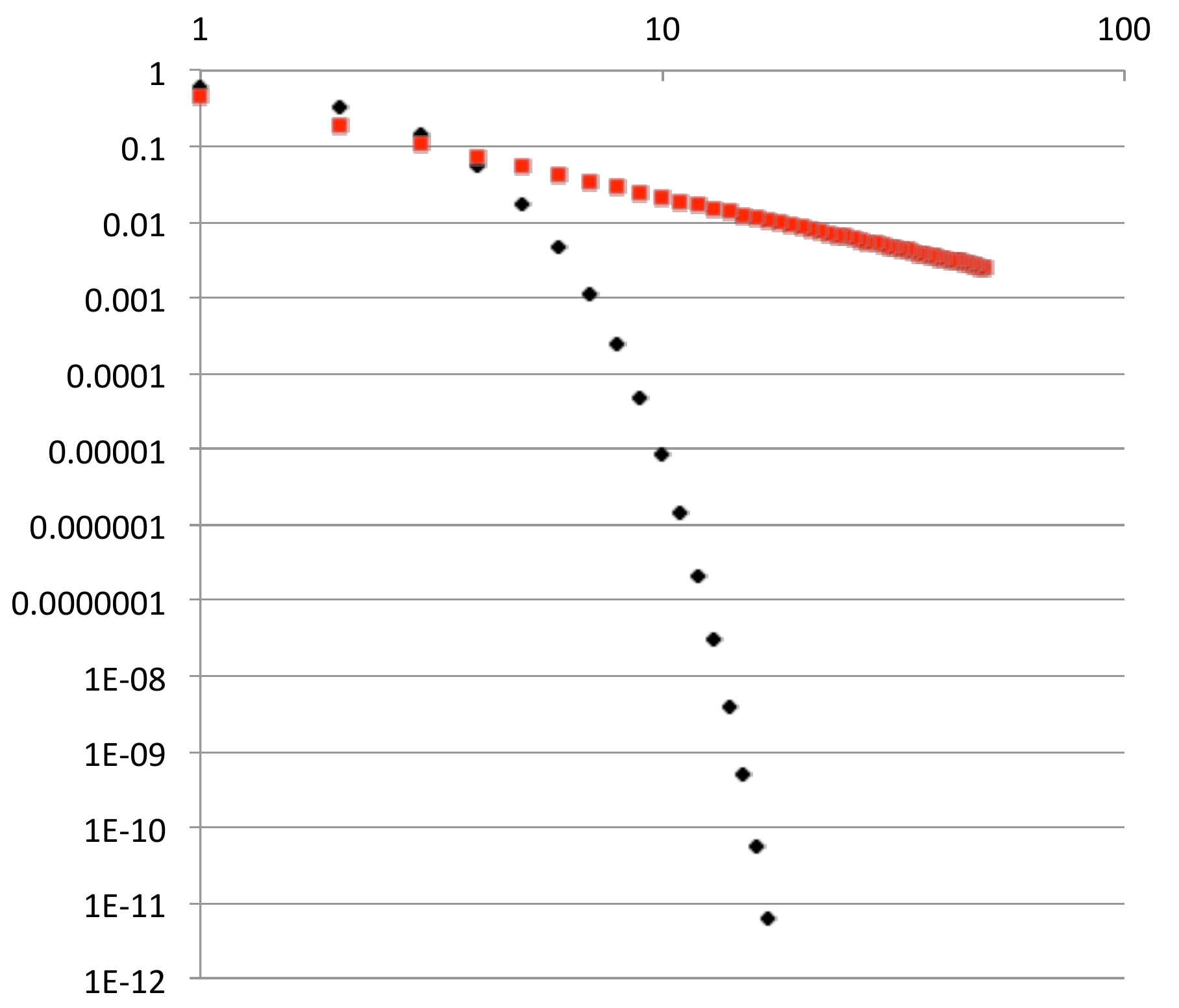}
\includegraphics[width=0.49\textwidth]{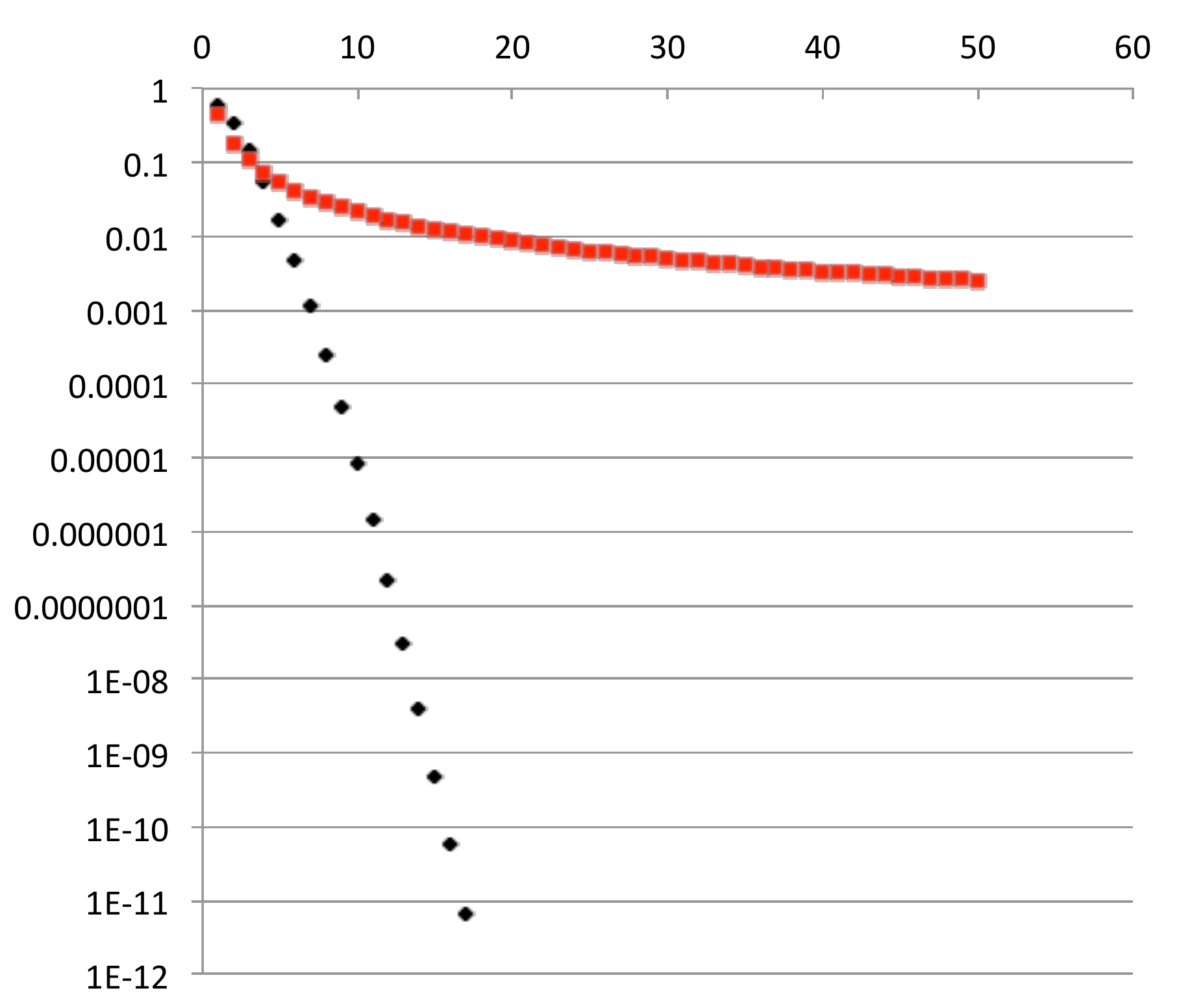}
 \caption{\label{fig:dist} Comparison of thin tailed Poisson distribution (black) with fat tailed power law distribution (red) that has a much higher probability of large events. Left panel uses logarithmic axes, right has logarithmic vertical axis only.}
\end{figure}

Empirical evidence for the existence of individuals that cause a much larger number of other cases than the average has been found in studies of the SARS epidemic \cite{SARS1,SARS2,SARS3}. Such incidents were reported in every region of the epidemic and after control measures were implemented and played an important role in the outbreak. Traditional studies consider such ``superspreader" individuals as anomalies rather than as indicating a fat tailed distribution with implications for evaluation of risks. 

The primary method used by public health authorities for containing an Ebola outbreak is contact tracing. In this method an individual who is infected is asked about people with whom he or she has been in contact during the contagious phase. They are then subject to a monitoring and isolation protocol for 21 days to prevent further spread. When there was one individual in Dallas that was sick, there were more than 100 potential contacts \cite{Dalascontacts}, many more if we count the people contacted for being on the plane in which an infected nurse flew \cite{nurseflight}. This does not include everyone who reported symptoms but did not have other reasons to be suspected and so were rapidly dismissed from consideration \cite{nonEbolacases}. If every outbreak has only a few individuals who are sick this can work. However, if there are 100, we might have to trace 10,000 people which becomes much less practical and perhaps impossible. 

Public health authorities are relying upon the assumption that even if a few people get sick, they will be able to contain the outbreak and, as they keep affirming, the public at large will not be at risk \cite{presidentatCDC,FriedenSept30}. All they have to do to achieve this is to identify enough contacts so that the resulting effective $R_0$ is less than one. A few people would get sick but there would not be a major outbreak as the number would decrease from generation to generation thereafter.

However, this assumes that a model with a specific value of $R_0$ describes what can be expected to happen in the US as it does in Africa, i.e. a thin tailed distribution holds.

The difference between situations in which thin tailed distributions and fat tailed distributions apply is whether the local events are essentially independent, or they are not independent, so they combine to create larger events.  

In subsistence agrarian societies in rural Africa, people act mostly in family groups. One person is not likely to be in close contact with much more than about 10 or 20 family members. If the probability of transmission for a single contact is low, then we end up with a few transmission events per person, which is what has been observed. The extent of transmission events by a single individual is bounded by the number of contacts they have. In urban areas in Africa and in the US, the nature of the contact network is different. 

The high density urban areas of West Africa, including Monrovia in Liberia, have people packed much closer together, and they have to do a variety of things that bring them in contact with each other to get food each day. Going to the market is one such activity. This can be expected to lead to a higher $R_0$ and a much larger outbreak, one that overwhelmed the public health efforts to contain it. We often hear about the problem with having too few hospital beds \cite{hospitalbeds}, but a key problem in urban areas of West Africa is that the main tool used by public health authorities to limit outbreaks is contact tracing. Contact tracing in a dense urban environment is much more difficult than in far rural areas or even less dense urban areas where Ebola has been present before. It becomes essentially impossible when there are many cases. Without contact tracing, it is hard if not impossible to contain an outbreak. While the contacts in Monrovia are of a higher frequency 
than those in rural areas, these are still poor societies. This makes people mostly connected locally to others in their neighborhood. Still, there are some ways in which they are connected more generally. Among the major reported sources of transmission is taking taxis to a hospital for care \cite{taxiliberia}. Taxis often carry multiple individuals going in the same direction and, even if not, a sick person leaving bodily fluids in the taxi can lead to transmission to subsequent passengers as well.

As society develops, the actions of people become connected in networks that are more interconnected and highly heterogeneous. Different people have very different levels of connection to others. Thus, the network of connections can be expected to be much more heterogeneous in the US. Individuals play varied roles in society and are connected to others to a different degree. Various ways that are used to characterize social networks show that they generally have fat tail distributions \cite{AlbertBarabasi}. We can infer that most people are not in a lot of direct physical contact with others, but some are. In a high density city the contact rate is much higher.

We can point to a few examples where transmission by direct contact might take place: a masseuse giving multiple massages every day, a politician that shakes many hands, and a nurse in a hospital (not one who is caring for Ebola patients) who goes from patient to patient. There are also events where many people are in direct contact including conferences in which people shake hands, and groups who go social dancing. Indirect contact through body fluids (including sweat as well as urine, vomit, blood and feces) deposited on a surface by one person in a place that other people touch it is a less probable way to transmit the disease, according to the experience in Africa, but it is still considered possible \cite{CDCinfect}. There are many such indirect contacts for restaurant or cafeteria workers, and in particular places, such as public buses and trains and public toilets, many people may follow where a single person has been. Unlike the dominant kinds of contact in families in poor societies, most of these contacts take place anonymously between people who do not know each other. This would make contact tracing or even knowing who is at risk much more difficult.

Let us translate this into a specific scenario. The doctor who returned from Guinea and went out into public places and on public transportation in New York and had symptoms the next day should be assumed to have a low probability of infecting anyone through those travels. But low probability does not mean no probability in this case. If there is one person who is infected, that person is not a high likelihood carrier. If he or she started to show symptoms the possibility of Ebola would be reasonably dismissed either by them or any health authority they contacted. A hospital would dismiss their having Ebola because they were not a traveler to west Africa. Hospitals do this every day for many cases where symptoms are similar to the early symptoms of Ebola, which are like those of the flu or other viral infections. CDC protocol explicitly states that such individuals should only receive routine medical evaluation and care \cite{CDC4}. Even if the symptoms got worse, an emergency room protocol would not consider them to have Ebola because of this. 

So let us say that an infected person happens to be a restaurant worker who is afraid of not showing up to work, or a masseuse that does not feel they can take off work because of prior appointments. Taking an Advil and going to work, perhaps by public transport, they are in indirect or direct contact over a few days with 10s or 100s of people. By that time perhaps the symptoms are so bad that they report to the hospital and someone realizes that Ebola should be tested for; at least it would be better if they do. Even so, we may have 10s of actual Ebola infections. At that point we cannot consider anybody in the entire city with fever as not being a possible Ebola exposed case. If the restaurant happens to be a popular place for travelers, say at one of the airports, we cannot treat anybody anywhere as if they are not a possible carrier. Currently we only have to treat fever as a possible symptom of Ebola if the person traveled to West Africa. This is a tiny fraction of all the cases of fever.

At this point the ability to use contact tracing to control the outbreak becomes doubtful and more severe actions must be taken including changes in behavior of the public. For healthcare facilities to provide intensive care to all the infected individuals, and attention to the large number of symptomatic but not Ebola infected individuals, may require compromising normal operations. How such an outbreak would eventually be contained is unclear as the real world experience is limited or non-existent. Surely it would not be easy to do.

What causes the fat tail distribution to arise in this case? How are the infection events dependent on each other? In a standard statistical treatment, we would average over all professions getting a number that is dominated by low contact workers, perhaps office workers sitting at their desk most of the day, who do not have much physical or indirect contact. In an imaginary world this would work if people switched professions every hour. Then the probability that a single person infected many others would be the same as any other and would equal the average. The reason that this is not the case is that a person does not switch profession from hour to hour. When someone has a profession that involves a lot of contact, if they are sick then there are many transmission events that are linked to each other. Thus the traditional statistical assumptions do not apply. 

Other, perhaps behavioral, scenarios might be considered, as in someone who is at an intermediate level of risk and self-monitoring has a few drinks (there are no restrictions on this in the CDC guidelines) and as a result acts in less responsible ways, violates guidelines for travel or proximity to others, and even might throw up perhaps just as he or she is beginning to show symptoms. Or someone becomes sick from food poisoning and throws up not because of Ebola but just as symptoms are developing.

These scenarios are only a few of many but they illustrate the problem of having a highly heterogeneous and interconnected network. The most likely scenario is that no one will be infected by the infected doctor's travels. The most likely scenario is that even if one person is infected they will not be one of the highly connected people and will infect only a few others that they are in contact with on a daily basis and know personally. This will give time to find them and isolate them. In a world of thin tailed distributions it is enough to identify the most likely scenario and any other scenario is going to be about the same. But in a world of fat tailed distributions there are other kinds of scenarios that are not the same and they can happen and often do.

The importance of uncertainty and risks that are associated with it is manifest in this discussion. Unlike the normal distributions commonly assumed in statistical approaches, fat tailed distributions lead to extreme events, i.e. much higher risks. This tends to lead to surprise. Indeed we have already seen such surprise in public health response in the US to Ebola, in the hospital response in Dallas and in the infections of nurses and the flight one of them took when she began to have symptoms. In the second installment of this series we will review in greater detail the uncertainty that arises because existing knowledge arose in rural West Africa, a different context from that of New York or other cities, or even rural areas of the US. One example of a relevant difference is the weather. Daily temperatures consistently reached highs in the upper 90s during the outbreak that served as the primary location of transmission analysis \cite{Dowell,Roels} and air conditioning is not a relevant factor there. Significantly lower room temperature may change the likelihood of indirect transmission through surfaces, an additional and potentially significant source of uncertainty about how outbreaks may spread in the US and other countries.

In conclusion, I note that the opinion of public health authorities is not being well received by the public \cite{public} and is being resisted by elected officials \cite{elected}. The mathematically based discussion of transmission given here supports a view that risks are much greater. Actions that reasonably anticipate risks are better than a reactive response. Since it can be expected that the actual number of infected individuals has a fat tail distribution, it is not impossible that an outbreak will overwhelm the ability of authorities to control it.

We thank John Sterman for helpful comments.


\begin{thebibliography}{999}

\bibitem{weiss} J. Weiss, CDC changes Ebola care guidelines for U.S. hospitals after Dallas case, Dallas News (Oct 13, 2014). \url{http://www.dallasnews.com/news/metro/20141013-cdc-changes-ebola-care-guidelines-for-hospitals.ece}
\bibitem{CDC1}
Safe management of patients with Ebola virus disease (EVD) in U.S. Hospitals, CDC (\url{http://www.cdc.gov/vhf/ebola/hcp/patient-management-us-hospitals.html})
\bibitem{CDC2}
Interim guidance about Ebola infection for airline crews, cleaning personnel, and cargo personnel, CDC (Updated October 15, 2014). \url{http://www.cdc.gov/quarantine/air/managing-sick-travelers/ebola-guidance-airlines.html}
\bibitem{CDC3}
Infection prevention and control recommendations for hospitalized patients with known or suspected Ebola virus disease in U.S. Hospitals, CDC (Updated October 20, 2014). \url{http://www.cdc.gov/vhf/ebola/hcp/infection-prevention-and-control-recommendations.html}
\bibitem{CDC4}
Interim U.S. Guidance for monitoring and movement of persons with potential Ebola virus exposure, CDC (Updated: October 27, 2014). \url{http://web.archive.org/web/20141028031921/http://www.cdc.gov/vhf/ebola/exposure/monitoring-and-movement-of-persons-with-exposure.html}
\bibitem{DHS}
Statement by Secretary Johnson on travel restrictions and protective measures to prevent the spread of Ebola to the United States, Department of Homeland Security, US Government (October 21, 2014). \url{http://www.dhs.gov/news/2014/10/21/statement-secretary-johnson-travel-restrictions-and-protective-measures-prevent}
\bibitem{States1}
H. Yan and G. Botelho, Ebola: Some U.S. states announce mandatory quarantines Ñ now what?, CNN (October 27, 2014). \url{http://www.cnn.com/2014/10/27/health/ebola-us-quarantine-controversy/index.html}
\bibitem{States2}
J. Ax and E. Wulfhorst, States stand firm on Ebola quarantines despite White House pressure, Reuters (October 27, 2014). \url{http://www.reuters.com/article/2014/10/27/us-health-ebola-usa-idUSKBN0IF03L20141027}
\bibitem{States3}
A. Sifferlin, N.Y. State relaxes Ebola quarantine rules, Time (October 26, 2014). \url{http://time.com/3540564/ny-state-loosens-quarantine-rules-for-ebola/}
\bibitem{Fauci}
B. Bell, Infectious disease specialist Dr. Anthony Fauci rejects mandatory quarantine, ABC News (October 26, 2014). \url{http://abcnews.go.com/Health/infectious-disease-specialist-dr-anthony-fauci-rejects-mandatory/story?id=26465651}
\bibitem{NEJM}
J. M. Drazen, R. Kanapathipillai, E. W. Campion, E. J. Rubin, S. M. Hammer, S. Morrissey, and L. R. Baden, Ebola and quarantine, New England Journal of Medicine. DOI: 10.1056/NEJMe1413139 (October 27, 2014).
\bibitem{Duncan1}
G. Botelho, M. Martinez, Frustrated woman quarantined with sheets, towels soiled by Ebola patient, CNN (October 3, 2014) \url{http://www.cnn.com/2014/10/02/us/texas-woman-quarantine-ebola-thomas-duncan/}
\bibitem{Duncan2}
Ebola patient's family legally quarantined after not complying with requests to stay home, AP (Oct 3, 2014)
\url{http://www.wccbcharlotte.com/news/top-stories/Ebola-Patients-Family-Legally-Quarantined-After-Not-Complying-With-Requests-To-Stay-Home-278020491.html}
\bibitem{Duncan3}
A. Park, HereÕs who Is being monitored for Ebola, Time Magazine (Oct. 16, 2014), \url{http://time.com/3513358/ebola-cdc-monitored/}
\bibitem{Duncan4}
K. Weintraub, Quarantine politics: {W}hy authorities push voluntary isolation in face of Ebola, National Geographic (Oct. 20, 2014), \url{http://news.nationalgeographic.com/news/2014/10/141020-ebola-quarantine-isolation-health-medicine/}
\bibitem{Duncan5}
L. Szabo, Ebola patient's family completes 21-day quarantine, USA Today (October 19, 2014) \url{http://www.usatoday.com/story/news/nation/2014/10/19/ebola-quarantine-ends/17443059/}
\bibitem{Dowell}
S.F. Dowell, R. Mukunu, T.G. Ksiazek, A.S. Khan, P.E. Rollin, and C.J. Peters, Transmission of Ebola hemorrhagic fever: A study of risk factors in family members, Kikwit, Democratic Republic of the Congo, 1995. Commission de Lutte contre les Epidemies a Kikwit. \emph{The Journal of Infectious Diseases}. \textbf{179} Suppl 1:S87-91 (February 1999).
\bibitem{Roels}
T.H. Roels, A.S. Bloom, J. Buffington, et al. Ebola hemorrhagic fever, Kikwit, Democratic Republic of the Congo, 1995: Risk factors for patients without a reported exposure. \emph{The Journal of Infectious Diseases}. \textbf{179} Suppl 1:S92-97 (February 1999).
\bibitem{Kersteins}
B. Kerstiens, F. Matthys, Interventions to control virus transmission during an outbreak of Ebola hemorrhagic fever: Experience from Kikwit, Democratic Republic of the Congo, 1995. \emph{The Journal of Infectious Diseases}. \textbf{179} Suppl 1:S263-267 (February 1999).
Ebola haemorrhagic fever in Sudan, 1976. Report of a WHO/International study team. Bulletin of the World Health Organization. 56(2):247-270 (1978).
\bibitem{Baron}
R.C. Baron, J.B. McCormick, and O.A. Zubeir, Ebola virus disease in southern Sudan: Hospital dissemination and intrafamilial spread. \emph{Bulletin of the World Health Organization}. \textbf{61}(6):997-1003 (1983).
\bibitem{Muyembe}
J.J. Muyembe-Tamfum, M. Kipasa, C. Kiyungu, and R. Colebunders, Ebola outbreak in Kikwit, Democratic Republic of the Congo: Discovery and control measures. \emph{The Journal of Infectious Diseases}. \textbf{179} Suppl 1:S259-262 (February 1999).
\bibitem{Francesconi}
P. Francesconi, Z. Yoti, S. Declich, et al. Ebola hemorrhagic fever transmission and risk factors of contacts, Uganda. \emph{Emerging Infectious Diseases}. \textbf{9}(11):1430-1437 (November 2003).
\bibitem{Wamala}
J. F. Wamala, L. Lukwago, M. Malimbo, et al. Ebola hemorrhagic fever associated with novel virus strain, Uganda, 2007-2008. \emph{Emerging Infectious Diseases}. \textbf{16}(7):1087-1092 (July 2010).
\bibitem{whoreport}
World Health Organization (WHO), Ebola response roadmap situation report, WHO (October 22, 2014). \url{http://apps.who.int/iris/bitstream/10665/137091/1/roadmapsitrep22Oct2014_eng.pdf?ua=1}
\bibitem{DCS}
Y.~Bar-Yam, Dynamics of Complex Systems, Westview Press (1997).
\bibitem{Meltzer}
M. I. Meltzer, C. Y. Atkins, S. Santibanez, B. Knust, B. W. Peterson, E. D. Ervin, S. T. Nichol, I. K. Damon, and M. L. Washington, Estimating the future number of cases in the Ebola epidemic - Liberia and Sierra Leone, 2014 - 2015, Morbidity and Mortality Weekly Report (MMWR), Centers for Disease Control and Prevention (September 26, 2014). \url{http://www.cdc.gov/mmwr/preview/mmwrhtml/su6303a1.htm}
\bibitem{R02}
Ebola virus disease in west Africa --- The first 9 months of the epidemic and forward projections. Report of WHO Ebola Response Team. \emph{New England Journals of  Medicine}. \textbf{371}:1481-1495. [DOI: 10.1056/NEJMoa1411100] (October 22, 2014).
\bibitem{R03}
C.L. Althaus. Estimating the reproduction number of Ebola virus (EBOV) During the 2014 Outbreak in West Africa. \emph{PLOS Currents Outbreaks}. Edition 1. [DOI: 10.1371/currents.outbreaks.91afb5e0f279e7f29e7056095255b288] (September 2, 2014).
\bibitem{dismiss1}
T. Randall, Five threats more terrifying than Ebola arriving in the U.S., Bloomberg (August 5, 2014). \url{http://www.bloomberg.com/news/2014-08-05/five-threats-more-terrifying-than-ebola-arriving-in-the-u-s-.html}
\bibitem{dismiss2}
J. Belluz, Ebola has up to an 18\% chance of coming to America. Here's why you do not need to panic. Vox (September 7, 2014). \url{http://www.vox.com/2014/9/6/6111275/chances-of-ebola-virus-spreading-to-america-on-flights-planes}
\bibitem{dismiss3}
P. Edelstein, The illness you really should worry about (and itÕs not Ebola), Elsevier Connect (October 30, 2014). \url{http://www.elsevier.com/connect/the-illness-you-really-should-worry-about-its-not-Ebola}
\bibitem{SARS1}
J. Wallinga and P. Teunis, Different epidemic curves for severe acute respiratory syndrome reveal similar impacts of control measures. \emph{American Journal of Epidemiology}. \textbf{160}(6):509-516. (September 15, 2004).
\bibitem{SARS2}
C. Dye and N. Gay, Modeling the SARS epidemic. \emph{Science}. \textbf{300}:1884-1885. (June 20, 2003).
\bibitem{SARS3}
S. Riley, C. Fraser, C.A. Donnelly, et al. Transmission dynamics of the etiological agent of SARS in Hong Kong: Impact of public health interventions. \emph{Science}. \textbf{300}: 1961-1966. (June 20, 2003).
\bibitem{Dalascontacts}
S. Levine, Up to 100 people being monitored in Dallas for Ebola, Politico (October 2, 2014). \url{http://www.politico.com/story/2014/10/ebola-us-texas-thomas-eric-duncan-111553.html}
\bibitem{nurseflight}
R.C. Haidet, Frontier expands Ebola notifications to 800 passengers, WKYC (October 17, 2014). \url{http://www.wkyc.com/story/news/health/2014/10/17/frontier-expands-ebola-notifications-to-800-passengers/17424245/}
\bibitem{nonEbolacases}
P. Stewart, Ebola scare at Pentagon after woman vomits in parking lot, Reuters (October 17, 2014). \url{http://www.reuters.com/article/2014/10/17/us-ebola-health-usa-pentagon-idUSKCN0I61X820141017}
\bibitem{presidentatCDC} Remarks by the President on the Ebola outbreak, Office of the Press Secretary, The White House (September 16, 2014). \url{http://www.whitehouse.gov/the-press-office/2014/09/16/remarks-president-ebola-outbreak}
\bibitem{FriedenSept30} T. Frieden, CDC Confirms First Ebola Case Diagnosed in United States, Centers for Disease Control and Prevention (CDC) (September 30, 2014). \url{http://www.cdc.gov/media/releases/2014/t0930-ebola-confirmed-case.html}
\bibitem{BlackSwan} 
N. N. Taleb, The Black Swan: The impact of the highly improbable fragility. Random House (2010)
\bibitem{hospitalbeds}
M. Fox, Ebola spreading 'Exponentially' as patients seek beds in Liberia, NBC News (September 8, 2104). \url{http://www.nbcnews.com/storyline/ebola-virus-outbreak/ebola-spreading-exponentially-patients-seek-beds-liberia-n198516}
\bibitem{taxiliberia}
Ebola situation in Liberia: Non-conventional interventions needed. WHO Situation Assessment (September 8, 2014). \url{http://www.who.int/mediacentre/news/ebola/8-september-2014/en/}
\bibitem{AlbertBarabasi}
R. Albert and A.L. Barab‡si. Statistical mechanics of complex networks. \emph{Reviews of Modern Physics} \textbf{74}, 47 (2002).
\bibitem{CDCinfect}
Review of human-to-human transmission of Ebola virus, CDC Report (October 29, 2014). \url{http://www.cdc.gov/vhf/ebola/transmission/human-transmission.html}
\bibitem{public}
A. Blake, Poll: 80 percent want Ebola quarantines, The Washington Post (October 29, 2014). \url{http://www.washingtonpost.com/blogs/the-fix/wp/2014/10/29/poll-80-percent-want-ebola-quarantines/}
\bibitem{elected}
B. Chappell, Christie defends quarantine and jabs At CDC over Ebola, NPR (October 26, 2014). \url{http://www.npr.org/blogs/thetwo-way/2014/10/26/359082611/christie-defends-quarantine-and-jabs-at-cdc-over-ebola}

\end{thebibliography}
\end{document}